\documentclass[a4paper,12pt, epsfig]{article}
\usepackage{epsfig, cite}
\usepackage{amssymb}
\usepackage{amsfonts}
\usepackage{xypic}
\usepackage[matrix,arrow,curve]{xy}

\newskip\humongous \humongous=0pt plus 1000pt minus 1000pt

\newif\ifdtup

\catcode`\@=11

\@addtoreset{equation}{section}
\def\theequation{\thesection.\arabic{equation}}

\def\@normalsize{\@setsize\normalsize{15pt}\xiipt\@xiipt
\abovedisplayskip 14pt plus3pt minus3pt%
\belowdisplayskip \abovedisplayskip
\abovedisplayshortskip \z@ plus3pt%
\belowdisplayshortskip 7pt plus3.5pt minus0pt}

\def\small{\@setsize\small{13.6pt}\xipt\@xipt
\abovedisplayskip 13pt plus3pt minus3pt%
\belowdisplayskip \abovedisplayskip
\abovedisplayshortskip \z@ plus3pt%
\belowdisplayshortskip 7pt plus3.5pt minus0pt
\def\@listi{\parsep 4.5pt plus 2pt minus 1pt
      \itemsep \parsep
      \topsep 9pt plus 3pt minus 3pt}}

\relax

\catcode`@=12

\catcode`\@=11

\def\section{\@startsection{section}{1}{\z@}{3.5ex plus 1ex minus
    .2ex}{2.3ex plus .2ex}{\large\bf}}

\def\thesection{\arabic{section}}
\def\thesubsection{\arabic{section}.\arabic{subsection}}

\def\appendix{\setcounter{section}{0}
  \def\thesection{Appendix \Alph{section}}
  \def\thesubsection{\Alph{section}.\arabic{subsection}}
  \def\theequation{\Alph{section}.\arabic{equation}}}

\def\SymBoxes#1#2#3#4{\newdimen\un@t \un@t#3%
\raisebox{#1}{\rule{#2\un@t}{#4}\hskip-#2\un@t
\@tempdimb\un@t \advance\@tempdimb by-#4\@tempcntb#2\relax%
\@whilenum{\@tempcntb>0}\do{
\rule{#4}{\un@t}\hskip\@tempdimb \advance\@tempcntb by\m@ne}%
\hskip-#2\un@t \rule[\un@t]{#2\un@t}{#4}%
\rule[\un@t]{#4}{#4}\hskip-#4
\rule{#4}{\un@t}}\hskip-#4}                

\begin{document}

\newcommand{\beq}{\begin{equation}}
\newcommand{\eeq}{\end{equation}}
\newcommand{\bea}{\begin{eqnarray}}
\newcommand{\eea}{\end{eqnarray}}
\newcommand{\beas}{\begin{eqnarray*}}
\newcommand{\eeas}{\end{eqnarray*}}
\newcommand{\defi}{\stackrel{\rm def}{=}}
\newcommand{\non}{\nonumber}
\newcommand{\bquo}{\begin{quote}}
\newcommand{\enqu}{\end{quote}}
\renewcommand{\(}{\begin{equation}}
\renewcommand{\)}{\end{equation}}
\def\de{\partial}
\def\Om{\ensuremath{\Omega}}
\def\Tr{ \hbox{\rm Tr}}
\def\H{ \hbox{\rm H}}
\def\HE{ \hbox{$\rm H^{even}$}}
\def\HO{ \hbox{$\rm H^{odd}$}}
\def\HEO{ \hbox{$\rm H^{even/odd}$}}
\def\HOE{ \hbox{$\rm H^{odd/even}$}}
\def\HHO{ \hbox{$\rm H_H^{odd}$}}
\def\HHEO{ \hbox{$\rm H_H^{even/odd}$}}
\def\HHOE{ \hbox{$\rm H_H^{odd/even}$}}
\def\K{ \hbox{\rm K}}
\def\Im{ \hbox{\rm Im}}
\def\Ker{ \hbox{\rm Ker}}
\def\const{\hbox {\rm const.}}
\def\o{\over}
\def\im{\hbox{\rm Im}}
\def\re{\hbox{\rm Re}}
\def\bra{\langle}\def\ket{\rangle}
\def\Arg{\hbox {\rm Arg}}
\def\Re{\hbox {\rm Re}}
\def\Im{\hbox {\rm Im}}
\def\exo{\hbox {\rm exp}}
\def\diag{\hbox{\rm diag}}
\def\longvert{{\rule[-2mm]{0.1mm}{7mm}}\,}
\def\a{\alpha}
\def\dag{{}^{\dagger}}
\def\tq{{\widetilde q}}
\def\p{{}^{\prime}}
\def\W{W}
\def\N{{\cal N}}
\def\half{\frac{1}{2}}

\def\del          {\partial}
\def\delbar       {\bar\partial}
\def\tr {\mbox{tr}}

\newcommand{\cF}{\cal F}
\newcommand{\cA}{\cal A}

\def\hsp{,\hspace{.7cm}}
\newcommand{\C}{\ensuremath{\mathbb C}}
\newcommand{\Z}{\ensuremath{\mathbb Z}}
\newcommand{\R}{\ensuremath{\mathbb R}}
\newcommand{\rp}{\ensuremath{\mathbb {RP}}}
\newcommand{\cp}{\ensuremath{\mathbb {CP}}}
\newcommand{\vac}{\ensuremath{|0\rangle}}
\newcommand{\vact}{\ensuremath{|00\rangle}                    }
\newcommand{\oc}{\ensuremath{\overline{c}}}
\newcommand{\wa}{\ensuremath{F^-_{(1,1)}}}
\newcommand{\ws}{\ensuremath{F^+_{(2,0)}}}
\begin{titlepage}
\rightline{\small SISSA 71/2008/EP}
\rightline{\small SPhT-T08/185}
\bigskip
\bigskip
\bigskip
\def\thefootnote{\fnsymbol{footnote}}

\begin{center}
{\large {\bf
Topology Change from (Heterotic) Narain T-Duality
  } }
\end{center}

\bigskip
\begin{center}
{\large  Jarah Evslin$^{1}$\footnote{\texttt{evslin@sissa.it}} and Ruben
Minasian$^{2}$\footnote{\texttt{ruben.minasian@cea.fr}}}
\end{center}

\renewcommand{\thefootnote}{\arabic{footnote}}

\begin{center}
\vspace{1em}
{\em  $^1${ SISSA,\\
Via Beirut 2-4,\\
I-34014, Trieste, Italy\\
\vskip .4cm
$^2$ Institut  de Physique Th\'eorique,
CEA/Saclay \\
91191 Gif-sur-Yvette Cedex, France  }
}
\end{center}

\vspace{.8cm}

\noindent
\begin{center} {\bf Abstract} \end{center}

\noindent

\noindent
We consider Narain T-duality on a nontrivially fibered $n$-torus bundle in the presence of a topologically nontrivial NS $H$ flux.  The action of the duality group on the topology and $H$ flux of the corresponding type II and heterotic string backgrounds is determined.   The topology change is specialized to the case of  supersymmetric $\mathbb{T}^2$-fibered torsional string backgrounds with nontrivial $H$ flux.  We prove that it preserves the global tadpole condition in the total space as well as on the base of the torus fibration. We find that some of these T-dualities exchange half of the field strength of an unbroken $U(1)$ gauge symmetry with the anti-selfdual part of the curvature of a physical circle fibration.  We verify that such T-dualities indeed exchange the supersymmetry condition for the circle bundle with that of the gauge bundle.

\vfill

\begin{flushleft}
{\today}
\end{flushleft}
\end{titlepage}

\hfill{}


\setcounter{footnote}{0}
\section{Introduction}

Dualities are one the most powerful and time-tested tools for finding new solutions to string theories.  Due to its perturbative nature, T-duality is the most reliable of the solution-generating dualities.  In the form most commonly used, that introduced by Buscher in Refs.~\cite{Buscher,B2}, it is a $\Z_2$ symmetry which applies to any trivial torus fibration in the presence of an exact NS 3-form field strength $H$ preserved by the torus action.
However it is well-known that the symmetry group can be larger than $\Z_2$ \cite{Narain}.  The symmetry of a $n$-dimensional torus fibration coupled to $r$ abelian vector fields may be as large as the Narain T-duality group $O(n,n+r;\Z)$.  

While the local action of the T-duality group has been known for some time \cite{MS}, recent developments in string compactifications have led to an interest in global aspects of T-duality. For example, one would like to understand the action of T-duality when the torus fiber degenerates. Such examples include  the  $\Z_2$ T-duality between an NS5-brane and a KK monopole with a nontrivial circle fibration, and also the SYZ picture of mirror symmetry \cite{SYZ}.  However  there are interesting and intricate problems associated with the topology change and obstructed T-duality even for nondegenerate torus fibrations.

In the case in which the circle fiber does not degenerate, the general topology change in type II under  $\Z_2$ T-dualities was understood in Refs.~\cite{BEM,BEM2}.   If the $\Z_2$ T-dualities apply even when the fibration and $H$ flux are nontrivial, one may then wonder whether the full $O(n,n+r;\Z)$ may be also extended to this case.  First one may ask whether there is a topological obstruction, independent of physical considerations, to the existence of the T-dual manifolds.  This was investigated in Ref.~\cite{MR} in the case $r=0$.  The authors found that an arbitrary $Gl(2n;\Z)$ transformation is obstructed, the unobstructed T-dualities are those which preserve the 4-class pulled back from the $k$-invariant of the base of the classifying space of Ref.~\cite{BS}.  They found that $O(n,n;\Z)$ transformations leave this class invariant, and so $O(n,n;\Z)$ T-dualities are unobstructed.


We will refer to the invariant class as the $Q_{(5)}$ charge and identify it with the NS5-brane charge in a type II compactification.  We will generalize it to a class $X_4$ in heterotic string theories where in a particular ansatz of supersymmetric compactifications it is equal to the tadpole and so must vanish. 

We are interested not just in maps that take topological spaces to other topological spaces, but in dualities of the full string theory that map solutions to other solutions.  Locally the full $O(n,n+r;\Z)$ symmetry does this, thus one must only check to see whether it preserves the topological conditions imposed by the physics.  In particular, one must check to see whether it is consistent with the heterotic tadpole condition in the presence of gauge fields.  We do this, and find that it is. We will find a number of obstructions to such T-dualities, which presumably relate them to nongeometric backgrounds as in Refs.~\cite{MR1,MR,Tfold}.  In particular, we will need to assume not only that the torus directions are isometries and that a number of vertical Lie derivatives vanish, but also that the gauge field strength be exact along the vertical directions.  In the ansatz that we will consider in Sec.~\ref{pressec}, this condition on the gauge bundle is imposed by supersymmetry.  In the case of a $\mathbb{T}^2$ fibration a sufficient condition for the elimination of global obstructions will be, as in the type II case, that the first Betti number of the base $B$ vanishes.

The geometric approaches to T-duality (see \cite{BEM, BEM2, MR, BS, MR1, Tfold, BHM} for a partial list) rely on a mathematical structure, known as a gerbe, which exists when the NS 3-form $H$ is closed.  This gerbe is used in the construction of a correspondence space, which is a torus fibration over spacetime where one simultaneously includes both the circle which is T-dualized and also its T-dual.  However the heterotic Bianchi identity implies that in general in heterotic string theory $H$ is not closed and so the familiar gerbe structure is not present\footnote{We will see in Subsec.~\ref{geosec} that instead of a 1-gerbe, the geometry is now described by a horizontal, trivial 2-gerbe on the total space of the gauge bundle together with a choice of trivialization.  This pair suffices to construct the correspondence space.}.  Nonetheless we will see that we are able to construct the correspondence space, and so we are still able to describe T-duality.

We are particularly interested in T-dualities that interchange a topologically nontrivial gauge bundle with a topologically nontrivial circle bundle in the compactification manifold.  The supersymmetric torsional heterotic backgrounds of Ref.~\cite{DRS,BY} on non-K\"ahler manifolds are obtained by fibering a two-torus over a $K3$ surface and have a non-trivial $H$-flux.  These provide a natural laboratory for studying such an exchange. Considering a subcase in which the unbroken gauge group contains an abelian component, we find that these Narain T-dualities exchange the anti-selfdual part of the circle bundle curvature with half of a $U(1)$ field strength.  This normalization is dictated by both the tadpole cancellation, and independently by the fact that the $U(1)$ gauge field must have an even Chern class for anomaly cancellation \cite{Freed,Distler}. We demonstrate that in this case T-duality exchanges the apparently different tree level supersymmetry conditions for the curvature and the gauge field strength.

We begin in Sec.~\ref{2sec} with a general description of Narain T-duality in type II and heterotic string theory, describing the characteristic classes $Q_{(5)}$ and $X_4$ and showing that the latter is indeed T-duality invariant. In Sec.~\ref{obsec} we discuss topological obstructions as well as corrections to T-duality when the connections have nontrivial vertical components.  This will allow us to describe the geometry that underlies our construction.  In Sec.~\ref{pressec} we describe the torsional heterotic backgrounds.  We fix a relative normalization of the geometrical and gauge fields, and show that with this normalization T-duality preserves the tadpole condition, as well as the supersymmetry condition and also the anomaly-canceling evenness of the gauge field strength.  In the appendix we give an example of a complicated topology-changing Narain T-duality in type II and show that it leaves the twisted K-theory of the total space invariant.

\section{T-duality  and $H$ flux } \label{2sec}


Let $X$ be a principal torus bundle with   fiber $\mathbb{T}^n$:
$$
\mathbb{T}^n\hookrightarrow X\stackrel{\pi}{\longrightarrow}B, 
$$
and denote the connection one-form by $\Theta$ and the fundamental vector field generating the torus action on $X$ by $K$. $\Theta$  takes values in $\mathfrak{t}:=\mathrm{Lie}\,\mathbb{T}^n\cong {\mathbb R}^n$, and $K\in \Gamma(TX\otimes \mathfrak{t}^*)$;  their contraction $\imath_K\,\Theta=1 \in \mathfrak{t}^*\otimes \mathfrak{t},$ and the Lie derivative $\mathcal{L}_K \Theta=0$. If the full string action respects the isometry, and $\mathcal{L}_K H =0$ as well, then one is led to the following decomposition of $H$ into horizontal and vertical parts:
\begin{equation}
\label{iso-H}
H = \pi^*  H_3 +  \pi^* H_{2,I}\wedge \Theta^I + \frac{1}{2}
\pi^* H_{1,IJ} \wedge \Theta^I \wedge \Theta^J  + \frac{1}{6}  \pi^* H_{0,IJK}\wedge \Theta^I \wedge \Theta^J \wedge \Theta^K, \label{hans}
\end{equation}
where $I,J = 1,...,n$.  
When $H$ is closed, it can be obtained from the action of exterior derivative on a local two-form
\beq
B = B_2 +  B_{1,I}\wedge \Theta^I + \frac{1}{2} B_{0,IJ} \wedge \Theta^I \wedge \Theta^J
\eeq
where the components $B_0$, $B_1$ and $B_2$ in general are not pullbacks.   The properties of the $H$ flux are essential for understanding global aspects and obstructions of T-duality. The fact that the condition 
$\mathcal{L}_K H =0$ for closed $H$ leads to the emergence of the dual torus bundle from the reduction of the gerbe structure has been the starting point of any formal treatment of T-duality for type II backgrounds. 

We will in general be interested in non-closed $H$, which will define the characteristic class
\beq
\label{x4}
X_4=\frac{dH}{4\pi^2}.
\eeq
While $X_4$ is exact in the total space $X$, the horizontal part of $X_4$ may be pulled back from a 4-form that is not necessarily exact on the base $B$, which by an abuse of notation we will also refer to as $X_4$.  One contribution to $X_4$ is the NS5-brane charge $Q_{(5)}$ wrapping the torus.  For nonabelian configurations, the contribution of  one unit of NS five-brane charge is equal to a single instanton \cite{DMW}.  In heterotic string theory there is a second contribution due to the Bianchi identity  
\beq
\label{bi-new}
dH = (2\pi)^2 X_4 = (2\pi)^2 \Bigl( Q_{(5)} -  \frac{\alpha'}{2} [p_1(TX)  - p_1(E)]  \Bigr) =  (2\pi)^2 Q_{(5)}  +  \frac{\alpha'}{4} [\tr R\wedge R -  \tr {\cF} \wedge {\cF}]  \,,
\eeq
where $p_1(TX)$ and  $p_1(E)$ are the Pontrjagin classes of the tangent and gauge bundles respectively.  Notice that the isometry condition implies that $p_1(TX)$ is horizontal. We will relate the nonhorizontalness of $p_1(E)$ with obstructions to T-duality in Sec.~\ref{obsec}.  Since the gauge invariance of the action requires that the right hand side of this expression is exact, it should integrate to zero on any four-cycle of $X$. While the base $B$ is not a cycle, in the compactifications that we will consider in Sec.~\ref{pressec} supersymmetry will imply that $X_4$ is horizontal and that its integral over the base also vanishes.   
\subsection{A warm up}

With our prime example in mind, we start with the case of a nontrivial principal 2-torus bundle $X$ over a simply-connected, compact base $B$ with second Betti number $b_2 >1$.  The circumference of each circle is set to $2\pi$.  The generalization to arbitrary rank torii and lower Betti numbers is straightforward.  The bundle is entirely characterized by the first Chern classes $c^1_1$ and $c^2_1$ of the two circles $S^1_1$ and $S^1_2$ of the torus.  There are large diffeomorphisms of the torus which mix these generators, forming a group which is a $\Z_2$ extension of $SL(2,\Z)$.  The lattice generated by the two Chern classes is invariant under this group, as are all topological invariants of $X$.  We will assume that the Chern classes are linearly independent and so this lattice is rank 2.  When the rank is nonmaximal then one combination of circles is trivially fibered.

In this section we will restrict our attention to $H$ fluxes with one leg along the torus.  In other words, we will only consider $H_2$ in the decomposition (\ref{hans}).  In later sections we will consider the effects of other components of $H$, finding that $H_3$ is T-duality invariant while $H_0$ and $H_1$ may lead to obstructions and invariably lead to corrections in the formulas for the T-dual geometry.

In the case of a trivial fibration with only $H_2$, the $H$ flux is an element of
\beq
H\in\H^2(B)\otimes \H^1(\mathbb{T}^2)=\Z^{2b_2}\subset \H^3(X). \label{trivco}
\eeq
In the case of a nontrivial fibration the generators of the torus cohomology in Eq.~(\ref{trivco}) are no longer closed.  Indeed,  the two connection one-forms  $\Theta^I = d\theta^I+A^I$  ($I=1,2$)   yield
\beq
d\Theta^I= \pi^*F^I  \, , \label{deq}
\eeq
making the cohomology of the total space of the torus bundle smaller than that of the trivial torus bundle.

More concretely, if $\beta^i$ form a basis of the second cohomology of $B$, where $i$ runs from $1$ to $b_2$, then a set of representatives of the allowed $H$ fluxes, up to exact forms, is
\beq
H= 2\pi \, h_{Ij}\Theta^I \cup\beta^j
\eeq
where the cup product is the graded-commutative multiplication in  cohomology;  on the level of differential forms it becomes the wedge product. The possible Chern classes for the $I$th principal circle bundles are
\beq
c_1^I = \frac{1}{2\pi} F^I =c^I{}_{j}\beta^j.
\eeq
Both $c_1^I$ and $h_{Ij}\beta^j$ have integral periods, and in fact may be lifted to integral cohomology.  While we will often treat them as differential forms, all of our results apply to the integral case as well.  In particular this integral lift leads to the possibility of new topological obstructions, and we will see that when the base $B$ has vanishing first Betti number no such obstructions appear.  The coefficients $h_{Ij}$ and $c^I{}_j$ are integers, and so the topological data may be summarized in an array of integers.

While $H$ is globally defined on the total space of the bundle, it is not necessarily globally defined on the base.  Its ill-definedness is characterized by the characteristic class $X_4 \in\H^4(B)$. We shall start with the type II theories, and thus the characteristic class $X_4$ is simply the NS5-brane charge $X_4=Q_{(5)}$ which  can be found using (\ref{deq})
\beq
Q_{(5)}= \frac{1}{(2\pi)^2} dH= \frac{1}{2\pi}  h_{Ij}d\Theta^I \cup\beta^j=h_{Ij}c^I{}_{k} \, \beta^k\cup\beta^j=h_{Ij}c^I{}_{k}n^{jk}_l z^l
\eeq
where $z^l$ is a basis of 4-cycles and $n^{jk}_l$ is the cup product multiplication table. 

We shall be mainly interested in orientable four-dimensional base spaces, whose fourth cohomology is isomorphic to the integers.  We will use this isomorphism to identify $Q_{(5)}$ with the integer 
\beq
Q_{(5)}= h_{Ij}c^I{}_{k}n^{jk} \label{5carica}
\eeq
where $n^{jk}$ is the intersection matrix on the 4-dimensional base $B$. 
Notice that the NS5-brane wraps the $\mathbb{T}^2$ fiber $Q_{(5)}$ times.  $Q_{(5)}$ times $\mathbb{T}^2$ is a homologically trivial cycle, which is natural as it is Poincar\'e dual to an exact form $dH$ on $X$, and so such an NS5-brane could decay.  However during the decay it would need to pass through configurations that are not invariant under the torus action, and so are out of our ansatz.  Therefore, if we restrict our attention to torus-invariant configurations, the NS5-brane charge on $B$ is conserved.  In the heterotic case this conservation will be more important as it will be dictated by supersymmetry and the tadpole condition.

The horizontality of $Q_{(5)}$ is a consequence of the fact that it is Poincar\'e dual to the vertical NS5-branes.  It implies that $i_K Q_{(5)}$ vanishes which is necessary for $i_K H$ to be closed and thus for T-duality to be defined.  If the NS5-branes were instead horizontal, then the T-dual would include KK monopoles on which the torus fiber degenerates.  In this paper we will not consider degenerating torii, which implies that our NS5-branes must remain vertical, as is implied by our condition that the Lie derivative of $H$ vanishes.

\subsection{Narain T-duality in type II}
How does Narain T-duality act in this framework?  For each two-cycle $\beta^j$ on the base, the relevant topological data consists of the integral four-vector
\beq
v_j=(c^1{}_{j},c^2{}_{j},h_{1j},h_{2j}).
\eeq
In other words the configuration is characterized by four elements of $\H^2(B)$.  In type II string theory Narain T-duality acts on these elements via $O(2,2;\Z)$ transformations.   Transformations with determinant $-1$ interchange type IIA and IIB while those with determinant $1$ preserve the theory.   

The left $O(2;\Z)$ subgroup rotates the left movers $L$ and the right $O(2;\Z)$ rotates the right movers $R$.  The torus and its T-dual correspond to the combinations $L+R$ and $L-R$ respectively.  Therefore in the basis in which elements $g$ of $O(2,2;\Z)$ satisfy 
\beq
g^\top \eta \,g= \eta
\label{conoluce}
\eeq
they preserve the metric $\eta$
\beq
\eta =\left(\begin{array}{cccc}0&0&1&0\\0&0&0&1\\1&0&0&0\\0&1&0&0\end{array} \right) \label{h}
\eeq
and the $4$ by $b_2$ matrix $(c^1{}_{j},c^2{}_{j},h_{1j},h_{2j})$ transforms in the fundamental of $SO(2,2;\Z)$, corresponding to matrix multiplication by $g$.  In other words the rotation on each generator of $\H^2(B)$ is performed separately, although they are all performed with the same $O(2,2;\Z)$ matrix
\beq
\left(
\begin{array}{c}
c^1{}_{j}\\{c}^2{}_{j}\\{h}_{1j}\\{h}_{2j}
\end{array}\right)
\stackrel{\rm{T}}{\longrightarrow}\left(
\begin{array}{c}
\hat{c}^1{}_{j}\\\hat{c}^2{}_{j}\\\hat{h}_{1j}\\\hat{h}_{2j}
\end{array}\right)
=
g\left(
\begin{array}{c}
c^1{}_{j}\\{c}^2{}_{j}\\{h}_{1j}\\{h}_{2j}
\end{array}\right).
\eeq
We will consider the action of such transformations on type II string theory compactifications.  These are locally just ordinary Narain T-dualities, and so it seems reasonable to conjecture that they are dualities of the string theory even when $c$ and $h$ are not exact, generalizing the strategy of Ref.~\cite{BEM}.  In this note we explore the consequences of such a conjecture.

What do Narain T-dualities do to the $Q_{(5)}$ charge?  The charge was found in (\ref{5carica}) to be $h_{Ij}c^I{}_{k}n^{jk}$.  This is just the inner product of the vectors $v_j$ and $v_k$ under the metric $\eta$ of Eq.~(\ref{h})
\beq
Q_{(5)}=n^{jk}(v_j,v_k)=n^{jk}v_j^\top \eta v_k
\eeq
and is therefore invariant under $O(2,2;\Z)$ Narain T-dualities
\beq
\hat{Q}_{\rm{5}}=n^{jk}(\hat{v}_j,\hat{v}_k)=n^{jk}\hat{v}_j^\top \eta \hat{v}_k=n^{jk}v_j^\top g^\top \eta g v_k = n^{jk}v_j^\top \eta v_k = Q_{(5)}.
\eeq
Therefore Narain T-duality preserves the $Q_{(5)}$ charge as in Ref.~\cite{MR}.  This will be important in the supersymmetric compactifications of heterotic string theory that we will discuss later, where tadpole cancellation together with lemma 10 of Fu and Yau \cite{FuYau} will demand that the charge vanish.  It would be interesting to determine whether supersymmetry demands a similar cancellation in supersymmetric compactifications of type II.  

When $B$ is of dimension greater than 4, then one needs to consider the $Q_{(5)}$ charge on every 4-cycle of $B$.  However this argument may be applied cycle by cycle to demonstrate that the charge is invariant.  In particular we may write $Q_{(5)}$ directly in terms of the cocycles
\begin{eqnarray}
Q_{(5)}&=&c_1(S^1_1)\wedge\int_{S^1_1}H+c_1(S^1_2)\wedge\int_{S^1_2}H=\frac{1}{2}\left|\left|\left(\begin{array}{c}c_1(S^1_1)\\c_1(S^1_2)\\\int_{S^1_1}H\\\int_{S^1_2}H\end{array}\right)\right|\right|^2\\&=&\frac{1}{2}\left(c_1(S^1_1),\ c_1(S^1_2),\ \int_{S^1_1}H,\ \int_{S^1_2}H\right)\left(\begin{array}{cccc}0&0&1&0\\0&0&0&1\\1&0&0&0\\0&1&0&0\end{array} \right)\left(\begin{array}{c}c_1(S^1_1)\\c_1(S^1_2)\\\int_{S^1_1}H\\\int_{S^1_2}H\end{array}\right)\nonumber
\end{eqnarray}
where 2-cocycles are multiplied using the cup product, and so the contraction with the intersection matrix $n_{jk}$ is automatic.  Strictly speaking the notation $\int_{S^1_k}$ applies only at the level of differential forms, in integral cohomology it should be interpreted as a pushforward $\pi^k_*$ via the projection map $\pi^k$ of the $S^1_k$ fibration.

\subsection{Heterotic Narain T-duality} \label{hetsec}

Heterotic string theories, in addition to the Chern classes of the $\mathbb{T}^2$ fibrations and the $H$ flux, come equipped with a gauge bundle $V$.  If the gauge bundle contains $r$ abelian factors, then the T-duality group is $O(2,2+r;\Z)$.  

The gauge bundle is not arbitrary, as anomaly cancellation demands that it must lift to an $E_8\times E_8$ or $Spin(32)/\Z_2$ gauge bundle in the ultraviolet.  With a mild restriction on the embedding of the gauge symmetry in $E_8\times E_8$ or $Spin(32)/\Z_2$, this implies that the sum of the first Chern classes of the $U(1)$ factors is even \cite{Distler}.  In the $E_8\times E_8$ case this is necessary for the ultraviolet $SO(16)$ gauge symmetry to have an associated spinor bundle which combines with its associated adjoint bundle to form an $E_8$ bundle.  In the $Spin(32)/\Z_2$ case it is necessary for the $spin$ lift of the $SO(32)$ bundle. In fact we need a slightly stronger condition than that the total first Chern class be even, we need that each first Chern class will be individually even.  However, given any set of integers that sums to an even number, there will always be an even number of odd numbers.  One can choose pairs of odd numbers, and change into a basis where one considers the sums and differences.  In the new basis all numbers will be even.  Furthermore any integral transformation from an all even basis results in another all even basis, and so in particular the evenness condition will be T-duality invariant.  Following Ref.~\cite{BY} we will consider embeddings of this type. For other embeddings, the normalization of the $U(1)$ gauge field strength below will have to be modified appropriately.

In the case $r=1$ T-duality consists of all matrices $g$ preserving     
\beq
 \eta =\left(\begin{array}{ccccc}0&0&1&0&0\\0&0&0&1&0\\1&0&0&0&0\\0&1&0&0&0\\0&0&0&0&\textrm{1}\end{array} \right) \label{h2}
\eeq
which act via
\beq
\left(
\begin{array}{c}
c^1{}_{j}\\{c}^2{}_{j}\\{h}_{1j}\\{h}_{2j}\\ \frac{1}{2\pi\sqrt{2}}{\cF}_j
\end{array}\right)
\stackrel{\rm{T}}{\longrightarrow}\left(
\begin{array}{c}
\hat{c}^1{}_{j}\\\hat{c}^2{}_{j}\\\hat{h}_{1j}\\\hat{h}_{2j}\\ \frac{1}{2\pi\sqrt{2}}\hat{\cF}_j
\end{array}\right)
=
g\left(
\begin{array}{c}
c^1{}_{j}\\{c}^2{}_{j}\\{h}_{1j}\\{h}_{2j}\\ \frac{1}{2\pi\sqrt{2}}{\cF}_j
\end{array}\right)\hsp g^\top \eta \,g= \eta  \label{th}
\eeq 
where $\frac{1}{2\pi}{\cF}_j$ is the Chern class of the $U(1)$ gauge bundle integrated over the 2-cycle $j$.  We will see in Sec.~\ref{pressec} that this peculiar normalization for the gauge field is necessary for T-duality to be unobstructed and to preserve the tadpole condition.  At first glance it may seem as though a column in which some entries are integers and others are irrational cannot be acted upon by an integral $SO(n,n+r;\Z)$ matrix.  However the integrality of the entries of this matrix depends on a choice of basis for the preserved matrix $\eta$.  With the choice of basis in Eq.~(\ref{h2}) the matrix elements are not integral, and so there is no inconsistency.  In Sec.~\ref{pressec} we will work in the basis in which the matrix elements are integral.

Clearly T-dualities can change the topology of the spacetime.  In heterotic string theories, ignoring 5-branes for the moment, the topology of spacetime is restricted by the condition that the gauge-invariant NS field strength $H$ satisfy the Bianchi identity
\beq
dH=\frac{\alpha'}{4} [\tr R\wedge R -  \tr \cF \wedge \cF]
\eeq
where $R$ and $\cF$ are the spacetime and gauge curvatures.  The gauge-invariance of $H$ implies that the left hand side is exact, and so the right hand side is exact.  Thus, at the level of cohomology, $\tr R\wedge R =  \tr \cF \wedge \cF$, in other words they must be equal when integrated over any 4-cycle.  

If T-duality is really a symmetry of the full string theory, it must preserve this topological condition.  Intuitively, if the torus bundle is nontrivial then the torus fiber is a boundary, and so $\tr R\wedge R$ and $\tr \cF \wedge \cF$ are supported on the base.  In fact, $\tr R\wedge R$ will automatically be horizontal if we demand that our torus action is an isometry.  The horizontalness of $F$ may similarly be imposed in our ansatz, although we will see below that in Ref.~\cite{BY} it arises as a requirement for supersymmetry.  Cycles of the base which are linear combinations of the Chern classes of the circle bundles are not cycles of the total space, since that would require a global section of a nontrivial bundle.  Therefore the ``embeddings'' of such cycles in the total space will have boundaries, and so when integrating $dH$ over these embeddings, Stokes' theorem dictates that the answer is not 0, but rather the integral of $H$ over the boundaries.  Thus the form of the tadpole condition changes to
\beq
\frac{\alpha'}{4} \int_D [\tr R\wedge R -\tr {\cF} \wedge {\cF}]=\int_{\partial D} H. \label{tad2}
\eeq
Other cycles, those of the base $B$ which lift to cycles of the total space of the torus bundle $X$, are not affected by the T-duality and so their tadpole conditions will be automatically invariant.  

Therefore one needs to check the invariance of tadpole conditions of the form (\ref{tad2}).  We have not attempted such an analysis in general.  In the sequel we will consider the non-K\"ahler heterotic compactifications of Refs.~\cite{BY}.  These are supersymmetric compactifications on torus bundles over Calabi-Yau's.  In these cases Fu and Yau have demonstrated \cite{FuYau} that the right hand side of Eq.~(\ref{tad2}) vanishes.  We will demonstrate that the T-duality transformation (\ref{th}) indeed leaves the tadpole condition invariant in such compactifications.


\section{Global obstructions for heterotic T-duality} \label{obsec}


When the gauge bundle data has some vertical legs, the action of the T-duality might be obstructed.  The obstruction is a generalization of an obstruction that is well-known in the type II context, and so we will now describe it in a unified framework.  

  There are two ways to construct a torus bundle.  One might start with a manifold $X$ with a free torus action and define a bundle by letting a projection map take each point in $X$ to its torus orbit in the space of orbits $B$.  Another approach is to begin with $B$, then add a circle bundle to produce $P^1$, then fiber a circle over $P^1$ to obtain $P^2$ and so on.  The Chern class of the first circle bundle is necessarily in $B$, and so $P^1$ is just a 1-torus bundle in the sense of the previous construction.  

The two approaches differ when one adds the second circle.  The second circle may have a nontrivial holonomy over the first circle.  In this case, the $\mathbb{T}^2$ metric is not the usual orthonormal metric.  It is this modified metric which is used in the T-duality.  In particular the curvature of the second circle fibration may have nontrivial components with a leg along the first circle.  If $B$ is not simply connected, or more precisely if its first Betti number is nonzero, then it is even possible to fiber the second circle so that its first Chern class is not a pullback of an element of $\H^2(B)$, in which case we say that it has a vertical leg.  When this happens $P^2$ is no longer a principal $\mathbb{T}^2$ fibration, because the transition functions for the second circle depend on the first circle coordinate and so the structure group of the bundle is, for example, $SL(2)$ instead of $U(1)^2$.  When one includes the third circle there is again a new kind of obstruction, which is that one may fiber the third circle over the first two circles.  In this case the fiber is not $\mathbb{T}^3$, but rather is a nilmanifold.  In the presence of this kind of obstruction we say that the Chern class has two vertical legs.

Thus we have seen three new phenomena in the recursive construction of a torus bundle.  First, there may be vertical Wilson lines which lead to a trivial vertical Chern class.  Second, the Chern class may have a single vertical leg, in which case the torus bundle becomes nonprincipal.  This is only possible when the first Betti number of $B$ is nonzero.  Finally, the Chern class may have components with two vertical legs, in which case the fiber is no longer a torus.  

In this note we are interested in T-duality, and so we must consider not only the torus fiber, but also the (T-)dual torus.  As T-duality exchanges the curvature with the $H$ flux \cite{BEM}, all of these obstructions will be encoded in the $H$-flux.  The first corresponds to an $H$-flux with some vertical components, but whose cohomology class has a single vertical leg, more precisely its pushforward by the projection map may be pulled back from $\H^2(B)$.  The second corresponds to an $H$-flux whose cohomology class has two vertical legs and one horizontal leg, so that its pushforward by one of the circles is of the form of a Chern class with a single horizontal and vertical leg.  Again, this is only possible when the first Betti number of the base is nonzero.  Finally the $H$ flux may have a component which is entirely supported in the cohomology of the torus, which only happens when the torus is at least three-dimensional, and so is not in the ansatz that we consider below.

In the heterotic theory, in addition to the metric and $H$ flux, one adds a gauge bundle on the total space $X$.  This gauge bundle, as is clear if it is realized as a Kaluza-Klein bundle, has the same obstruction structure as the Chern class of a circle bundle and as the $H$ flux.  If the gauge bundle is entirely pulled back from the base, then one may use without modification the results of the previous section.  If it deviates from this by something exact, then the formulas are corrected as in Ref.~\cite{MS} but the topological considerations are as above.  We will give a geometric interpretation of these corrections shortly.  If on the other hand the cohomology class of the Chern class is not horizontal, then again T-duality will be obstructed.  This will never occur in the supersymmetric cases of Ref.~\cite{BY}, where even as a differential form the curvature in the vertical directions is equal to zero, but may occur in principle.

In every case one may construct an auxiliary space typically called the correspondence space. This is a fibration whose base is $B$ and whose fibers consist of the torus, the dual torus, and in the heterotic case also the gauge bundle $U(1)^r$.  In general this correspondence space may be constructed by fibering one circle at a time, and so one obtains the above obstructions and the fibers may not even be torii.  If instead it is fibered all at once, then T-duality acts as in the previous subsections.  If one deforms the configuration by exact curvatures, $H$ flux and field strengths, then the topology is not changed and so there are no obstructions.  However the metric on the torus changes.  For example, consider a single circle fiber and a single $U(1)$ gauge bundle.  If the $U(1)$ gauge bundle has a constant Wilson line along the circle, then the correspondence space will have a $\mathbb{T}^2$ fibration whose fibers have a complex structure determined by the Wilson line.  This means that the Wilson lines deform the metric of the correspondence space, and it is the horizontal metric of the correspondence space which is used in the T-duality maps of Refs.~\cite{MS,ST}.  In the supersymmetric case, at least in the ansatz of Ref.~\cite{BY}, the gauge field strength has no components with vertical legs and so there are no obstructions, however such Wilson lines are allowed.

The second and third obstructions in type II compactifications have been investigated in Refs.~\cite{Tfold,MR1,aussies,BHM}, where it is claimed that they are dual to nongeometric compactifications.  We are now in a position to describe their effects on the correspondence space of the heterotic theory.  The results are summarized in Table~\ref{tabella}.  In the case of the second obstruction, in other words when the $H$ flux has two vertical legs or the gauge field has one, the correspondence space is a nonprincipal torus bundle whose transition functions are special linear transformations of the torus.  The third obstruction corresponds to an $H$ flux component with 3 vertical legs or a gauge field strength with two.  In this case one of the circle fibers in the correspondence space is fibered over a torus which is also in the fiber with a Chern class given by the gauge field strength or the pushforward of the $H$ flux.  Therefore the total space of the fiber in the correspondence space is no longer a torus, it is a nilmanifold. 

The novelty for the analysis of the heterotic case is that $H$ is no longer closed.  If T-duality mixes $h=\int_{S^1} H$ with cohomology classes then $\int_{S^1} H$ needs to be closed.  This occurs when the field strengths $\cF$ have no vertical components.  The latter condition is satisfied, for example, if our gauge bundle on $X$ is pulled back from a gauge bundle on $B$, which then may be tensored with a flat bundle on the torus fibers.  As we have just mentioned, in the ansatz that we will consider below, supersymmetry demands that the gauge bundle be of this form \cite{BY}.  Such a flat bundle will not affect the topology of the configuration and so will not affect our topological tadpole arguments, but does affect the T-duality at the level of differential forms and so should be included in an analysis of the supersymmetry.  We will see in the next subsection that sometimes T-duality may be defined even when $\cF$ is not purely horizontal, so long as the vertical part is exact.

\subsection{Obstructions from nonhorizontal characteristic classes} \label{obsec*}

The $SO(n,n+r;\Z)$ T-duality group interchanges integral classes, in particular it exchanges horizontal closed forms.  However if the connections $\cA$ or $B$  or the curvatures $\cF$ or $H$ have vertical components, or if we are in the heterotic theory, the most obvious forms that we might like to exchange are no longer horizontal and closed.  In this subsection we will determine just when we are able to define horizontal closed forms upon which to act and we will construct these forms.

Let us return for a moment to type II and briefly review the obstructions following the discussion of \cite{BHM}.  As before we consider a manifold $X$ which is a principal $\mathbb{T}^n$ fibration and an $H$ flux which respects the isometries of the metric, i.e. satisfies $\mathcal{L}_K H =0$ and thus is of the form (\ref{iso-H}).

There are two obstructions to geometrical T-duality. We shall take the first one, namely $H_0 =0,$ for granted here (it boils down to imposing that not simply $H$ but also $B$ is annihilated by Lie derivative with respect to isometry generators $K$), and would like to re-examine the second - $H^1(B) = 0$, which is true for example if the base is simply connected. Due to $\mathcal{L}_K H =0$, it follows immediately that by contracting $H$ with the vector $K$ a closed two-form with integral periods can be constructed $F_{\#} = \imath_{K}H$, which can be viewed as a curvature of a $\mathbb{T}_{\#}^n$ bundle over $X$. In general, this is not yet the dual torus bundle. The latter should be freely exchangeable with $\mathbb{T}^n$ and thus unobstructed T-duality requires that the dual bundle sits directly over the base $B$. In other words $F_{\#}$ should be a horizontal form. It is not hard to calculate 
\begin{equation}
(F_{\#})_I :=\imath_{K_I}H=H_{2,I} -H_{1,IJ} \wedge \Theta^J \, , \qquad p^*(F_{\#}) = d \Theta_{\#}
\label{Fhash}
\end{equation}
and is not horizontal (as it contains $\Theta$). Here $p$ is the projection of the $\mathbb{T}_{\#}^n$ bundle to $X$.

However the curvature should be a pullback of some cohomology class on $B$, and thus by adding an exact piece one arrives at 
\begin{equation}
[(F_{\#})_I]_{dR}=\pi^*[H_{2,I}+B_{0,IJ}F^J]_{dR} = [\tilde{F}]_{dR} \label{ftwo}
\end{equation}
provided $B_0$ is globally defined. So if the base manifold $B$ is simply connected, there are no topological obstructions to treating the torus bundle obtained by geometrizing $H$ as a bundle over $B$.

The only subtlety is in the choice of the connection on the dual torus to be exchanged with $\Theta$:
\begin{equation}
\tilde{\Theta}_I=(\Theta_{\#})_I + B_{0,IJ} \Theta^J \, , \qquad d{\tilde \Theta} = {\tilde \pi}^*({\tilde F}) \, ,
\label{tTh}
\end{equation}
where ${\tilde \pi}$ is the projection form the dual torus bundle to $B$. Naively this expression, as well as Eq.~(\ref{ftwo}), is not invariant under large gauge transformations, which globally shift $B_0$ by an integer.  However this corresponds to an integral special linear transformation on the correspondence space and so leaves it invariant.  This implies that it does not affect the T-duality.  For example, after a $\Z_2$ T-duality on $S^1_I$ it would result in an integral special linear transformation on the dual Chern classes and so on the two torus fibers, which leaves the T-dual geometry invariant.  Note that the characteristic classes computed from $F_{\#}$ and from ${\tilde F}$ are the same.

We are now ready to go back to the heterotic string.  Let us for simplicity consider a single $U(1)$ gauge field ${\cal A}$ with a field strength ${\cal F}$. We shall restrict ourselves the same geometrical set-up as above, namely we have $H^1(B)=0$ and we assume that both $\mathcal{L}_K{\cal F} = 0$ and $\mathcal{L}_K{\cal A} = 0$. Hence we get 
\bea
{\cal A} &=& {\cal A}_1 + a_I \Theta^I \nonumber \\
{\cal F} &=& {\cal F}_2 + {\cal F}_{1,I} \wedge \Theta^I = (d {\cal A}_1 + a_I F^I) +  d a_I \Theta^I \, 
\eea
(once more - we are assuming that the Wilson line $a_I$ has no dependence on the torus coordinates). 

Let us now consider the  Bianchi identity $dH = \frac{\alpha'}{4}[ \tr R\wedge R - {\cal F} \wedge {\cal F}]$.  When decomposed into horizontal and vertical parts, this yields
\bea
\label{bi3}
d H_3 + H_{2,I} \wedge F^I &=& \frac{\alpha'}{4} [\tr R \wedge R - {\cal F}_2 \wedge {\cal F}_2]\nonumber\\
dH_{2,I} + H_{1,IJ} \wedge F^J &=& - \frac{\alpha'}{2}{\cal F}_2 \wedge {\cal F}_{1,I}\nonumber\\
dH_{1,IJ} &=&   \frac{\alpha'}{2}{\cal F}_{1,[I} \wedge {\cal F}_{1,J]}.
\eea

From the other side, using  $\mathcal{L}_K H = 0$, 
\beq
d\imath_{K_I} H= - \imath_{K_I} dH = - \frac{\alpha'}{2} d ({\cal A}_1 + a_J \Theta^J) \wedge  {\cal F}_{1,I},
\eeq
where we have used the fact that the circle action is an isometry and thus $i_K \tr(R^2)$ vanishes.
It is not hard now to see which closed two-form can serve as the curvature of the heterotic $\mathbb{T}_{\#}^n$ bundle:
\beq
(F_{\#})_I = \imath_{K_I} H + \frac{\alpha'}{2} a_I  {\cal F}  = (H_{2,I} +  \frac{\alpha'}{2} a_I {\cal F}_2) -(H_{1,IJ} - \frac{\alpha'}{2} a_{[I}  da_{J]} ) \wedge \Theta^J \,. \label{fhet}
\eeq
Since the curvature $(F_{\#})$ is not necessarily  horizontal, the $\mathbb{T}_{\#}^n$ bundle is not a pullback from $B$. It follows from the last equation of (\ref{bi3}) that the coefficient of the term proportional to $\Theta^J$ is closed. If $B$ has vanishing first Betti number and thus $B_{0,IJ}$ and $a_I$ are well defined functions, it is also exact. Thus we can define
\beq
H_{1,IJ} - \frac{\alpha'}{2} a_{[I}  da_{J]}  = d \mathcal{B}_{0,IJ} \, ,
\eeq
and a new connection $\tilde{\Theta}^J$ whose curvature is horizontal and is in the same cohomology class as $(F_{\#})$
\begin{equation}
[(F_{\#})_I]_{dR}=\pi^*[H_{2,I} +  \frac{\alpha'}{2} a_I {\cal F}_2 +\mathcal{B}_{0,IJ}F^J]_{dR} = [\tilde{F}]_{dR}. 
\label{ftwoX}
\end{equation}
The trio of connections $\left(\Theta^I, (\Theta_{\#})_I + \mathcal{B}_{0,IJ}\Theta^J, {\cal A} - a_I \Theta^I  \right)$ is acted upon by the $O(n,n+r;\Z)$ duality group in the fundamental (c.f. \cite{MS}).  When the base $B$ has a nonvanishing first Betti number (but $B_{0,IJ}$ and $a_I$ are still independent on the torus coordinates), this action will be obstructed in those configurations for which $\mathcal{B}_{0,IJ}$ or $a_I$ cannot be globally defined.  One may hope that, by adapting the generalized geometry to heterotic strings, it will be possible to perform the obstructed T-duality along the lines of \cite{GMPW}. We shall not pursue this here. Instead, we shall examine more closely an example of a supersymmetric background without Wilson lines.

Notice the similarity between the formulas for $F_{\#}$ in Eqs.~(\ref{ftwo}) and (\ref{fhet}) in type II and in the heterotic theory respectively.  To pass from type II to the heterotic theory, one need only replace $H$ by $H-a{\cal F}$.  In fact, both expressions are the closed form $dB$.  This may seem puzzling, because $dB$ is not gauge-invariant, $H=dB+a{\cal{F}}$ is gauge-invariant.  Any attempt at a definition of $dB$ from $H$ would fail to be well-defined if one considers a circle-valued family of configurations with a nontrivial instanton number.  However the assumption that $\cF$ is horizontal implies that the instanton charge is also horizontal and so is killed by the pushforward.  Therefore the pushforward of $dB$ may be globally well-defined.  This is essential, because the well-definedness of the fundamental string partition function implies that not $H$ but $dB$ is quantized, and so both T-dual field strengths $F_{\#}$ will lift to integral cohomology classes.  

\noindent
\begin{table}
\begin{tabular}{c|c|c}
\bf{Property of $(\pi_*H,F,\cal F)$}&\bf{Correspondence Space}&\bf{T-dual geometry}\\\hline
$a_I$ or $B_0$ nonzero&nonorthogonal torus&use nonorthogonal metric\\
\hline
$b^1(B)\neq 0$, one vertical leg&nonprincipal bundle&nongeometric \cite{MR,Tfold}\\
\hline
two vertical legs&fiber is not a torus&nongeometric \cite{Tfold,aussies}
\end{tabular}
\caption{{\textit{The three curvatures, $\pi_*H,\ F$ and $\cal F$ and their connections all appear on equal footing in the correspondence space.  If all are horizontal, then T-duality simply rotates them amongst each other.  More generally they may be intertwined.  This table illustrates the three levels of possible intertwinings in order of increasing complexity.  First, one may include a nontrivial connection $a_I$ or $B_0$ on the torus.  This does not prevent T-duality, but the torus in the correspondence space is nonorthogonal and its' metric appears in the Buscher rules.  In the next two levels the curvature has one or two vertical components as an element of cohomology, or the $H$ flux has two or three.  In these cases a geometric T-duality is obstructed.}}} \label{tabella}
\end{table}

\subsection{Heterotic geometry} \label{geosec}

Now that we have defined all of the closed forms that we need, we may provide a geometrical interpretation of our construction, which will be summarized in figure \ref{xy}.  In type II the usual geometrical construction of T-duality is that one defines a 1-gerbe on $X$ whose curvature is the closed 3-form $H$, and builds a correspondence space by fibering a dual $\mathbb{T}^n$ bundle over $B$ whose Chern classes are the pushforwards of $H$ via the $n$ projection maps $\pi_I$ of the original circle bundle.  The pushforwards live on a $\mathbb{T}^{n-1}$ bundle over $B$ whereas the Chern classes live on $B$, but if the pushforwards are horizontal then they define classes in the cohomology of $B$ and so the T-duality is unobstructed.

We claim that in the heterotic case we should consider the pullback $\mathcal{H}$ of $H$ to the total space of the gauge bundle $\rho:\mathcal P\rightarrow X$, or more precisely
\beq
\mathcal{H}=\rho^*H+\frac{\alpha\p}{4} \tr \cF\wedge {\mathcal{A}} \,,
\eeq
where ${\mathcal{A}}$ is the connection of the gauge bundle.  The trace is just a sum over the $U(1)$ factors of the unbroken abelian gauge group.  By construction, its dimensional reduction yields $H$ on $X$.  This $\mathcal{H}$ has the useful property:

\noindent
{\bf{Proposition: }}{\it{If $K$ is any constant vector field along the torus fiber, then $i_K \mathcal{H}$ is closed.}}

The connection $ \mathcal{A}$ is globally defined on $\mathcal P$, unlike $X$, and so $\mathcal{H}$ is globally defined.  In the exterior derivative of $\mathcal{H}$ the tr$\mathcal{F}\wedge\mathcal{F}$ terms cancel and one is left with tr$(R\wedge R)$, which is horizontal.  The Lie derivative of $\mathcal{H}$ with respect to $K$ vanishes, and so
\beq
d(i_K \mathcal{H})= - i_K d\mathcal{H}= - \frac{\alpha'}{4}i_K {\rm{tr}}R\wedge R=0
\eeq
where the last equality follows from the horizontality of $R$ in the gauge bundle.  Thus we have demonstrated the proposition.

The fact that $\mathcal{H}$ itself is not closed implies that there is not really a gerbe on $\mathcal P$.\footnote{Note that  the characteristic class $X_4$ defined in (\ref{x4}) gives rise to a closed 3-form $\imath_K X_4$ and  thus there is an associated gerbe structure.}  However only the closure of $i_K\mathcal{H}$ is necessary to define the correspondence space.  Once one has established that this is closed, then there is the problem of horizontality, which is solved by adding an exact piece as in the previous subsection.  Therefore while we do not quite have a 1-gerbe on the total space $\mathcal P$ of the gauge bundle, we are nonetheless able to construct the correspondence space because all of the pushforwards of $H$ are closed.

\begin{figure} \begin{equation} \label{correspondenceb}
\xymatrix @=3pc 
{ &(\mathcal H,\rho^*\pi^*\mathcal G)\ar[ddl]_{\pi\circ\rho}\ar[d]^{\pi_I}\ar[dr]&\\
&({\pi_I}_*{\mathcal H},0)\ar@{-->}[dd]^{\rm Subsec. \,\,\ref{obsec*}}&{\mathcal P}\ar[d]_\rho&U(1)^r\ar[l]\\
{\mathcal G}\ar[dr]&&X\ar[dl]^{\pi}&{\mathbb T}^n\ar[l]\\
&B&}  
\end{equation}
\caption{\textit{The Pontrjagin class of $B$ defines a 2-gerbe, whose pullback to the total space $\mathcal P$ of the gauge bundle is trivial and horizontal but not flat.  The NS field strength defines a trivialization of this gerbe, $\mathcal H$.  The horizontal 2-gerbe together with its trivialization suffice to construct the correspondence space because, using the proposition, it pushes forward to a flat 1-gerbe with a trivialization.  This trivialization defines a circle bundle on the image of the push forward.  If this circle bundle may be pulled back from $B$, then T-duality is unobstructed and it may be used to construct the correspondence space. }}\label{xy}
\end{figure}

While $\mathcal H$ does not quite define a 1-gerbe, it does define a trivialization of a trivial 2-gerbe with characteristic class Tr$(R^2)$.  This 2-gerbe is trivial by the Bianchi identity, but it is not flat, as the curvature is nonzero.  In fact it is the pullback of a 2-gerbe $\mathcal G$ from $B$, which need not be trivial.  We claim that the pair $({\mathcal H},\pi^*\rho^*{\mathcal G})$ of a trivial horizontal 2-gerbe and its trivialization is sufficient to construct the correspondence space.  It is sufficient because the pushforward of the 2-gerbe under any of the circle projection maps $\pi_I$ is a flat 1-gerbe, and so the pushforward of the trivialization is just a circle bundle.  This new circle bundle lives on the gauge bundle quotiented by a circle $S^1_I$; to construct the correspondence space we instead want a circle bundle on $B$.  The construction of the circle bundle on $B$ is then described by the procedure in Subsec. {\ref{obsec*}}.

\section{Torsional (non-K\"ahler) geometries} \label{pressec}

It has been shown recently that six-dimensional manifolds that are holomorphic $\mathbb{T}^2$ bundles over a $K3$ surface are consistent smooth backgrounds of heterotic string theory. Since there is an action of the $O(2,18;\Z)$ Narain lattice, this setup provides a natural laboratory for applying some of the wisdom we have acquired.

\subsection{The solution}

A few words about the background.  The metric on the six-manifold $X$ is \cite{DRS, GP}
\beq
\label{metric}
ds^2 = ds_B^2 + |d \theta + A|^2
\eeq
where $\theta = \theta^1 + i \theta^2$ and $A=A^1+iA^2$ are respectively the coordinate and the $U(1)$ potential  on $\mathbb{T}^2$. $\Theta = d \theta + A$ is a smooth connection one-form.

Provided that
\beq
F = dA  = F^+_{(2,0)} + F^-_{(1,1)} \in H^{2,+}(B) \oplus H^{2,-}(B)  \label{F}
\eeq
the metric is complex (the holomorphic three-form $\Omega = \Omega_B \wedge \Theta$ is conformally closed.) The natural hermitian (1,1) form
\beq
\label{sympl}
J = e^{2\phi} J_B + \frac{i}{2} \Theta \wedge \bar{\Theta} \label{J}
\eeq
is not closed. It is required by supersymmetry to be related to the $H$ flux via \cite{str,hull}
\beq
H=i(\delbar-\del)J
\eeq
which implies
\beq
\label{JH}
2i \del \delbar J = dH = \frac{\alpha'}{4} [\tr R \wedge R - \tr {\cal F} \wedge {\cal F} ] \, ,
\eeq
where $R$ and ${\cal F}$ are respectively the curvatures of the tangent bundle of $X$ and of the gauge bundle $E$. One particular consequence of this equation is that the right hand side should vanish when integrated over any closed four-dimensional submanifold.

We may now determine whether $h=\int_{S^1} H$ is closed.  The integral over the circle fiber may be written as an interior product with respect to a vertical vector $h=\imath_KH$.  Then the exterior derivative of $h$ is
\beq
dh=d\imath_K H=\mathcal{L}_K H - \imath_K dH=0-\imath_K \tr R^2 + \imath_K \tr {\cal F}^2=0
\eeq
where the fact that the circle action is an isometry implied that $i_K \tr(R^2)$ vanishes. The  stable gauge bundle required by supersymmetry is obtained, up to tensoring with a flat bundle, by pulling back to $X$ a stable bundle on the base $B$, and has no non-trivial dependence on the torus directions. Thus $\imath_K \tr({\cF}^2)$ vanishes due to horizontality of ${\cal F}$.  Therefore, as desired, $h$ defines a class in $\H^2(B)$.

\subsection{Compatibility of T-duality with  supersymmetry}

The $SO(2,18;\Z)$ heterotic Narain T-duality group includes dualities which exchange the Chern class of a $U(1)$ gauge bundle with the right-mover combination $h-c$.  Here $h$ is the pushforward $h=\pi_*H$ of the $H$ flux under the projection map of a circle fibration and $c$ is the Chern class of the same fibration.  The consistency of such an operation, or at least its compatibility with tree level supersymmetry, is nontrivial because the supersymmetry conditions are very different on both sides of this correspondence. 

The $U(1)$ gauge field needs to be a solution to the Hermitian Yang-Mills equations.  In other words, its field strength ${\cF}$ needs to satisfy
\beq
{\cF}^{(2,0)}={\cF}^{(0,2)}={\cF}_{ab}J^{ab}=0.
\eeq
On the other hand, the curvature of the circle bundle $F$, defined in Eq.~(\ref{F}) is much less constrained.  It only needs to be primitive on the base
\beq
F\wedge J_B=0. \label{prim}
\eeq
Thus naively the condition on circle bundles is less strict then that for $U(1)$ gauge bundles, and so they cannot be dual.  However it is only the right-moving part of the circle bundle which is exchanged with the gauge field, and this has a correction from the $H$ flux, which is
\beq
H=i(\overline{\partial}-\partial)J. \label{hdj}
\eeq
Thus it is not the supersymmetry conditions on $c$ and $F$ that need to agree, but rather those on $h-c$ and $F$.

To see this more explicitly, let us rotate Eqs.~(\ref{h2}) and (\ref{th}) by 45 degrees, so that the $SO(2,3;\Z)$ metric is diagonal $\eta = \textrm{diag (-1,-1,1,1,1)}$,
and the T-duality action is
\beq
\left(
\begin{array}{c}
{h}_{1j}+c_{1j}\\{h}_{2j}+c_{2j}\\{h}_{1j}-c_{1j}\\{h}_{2j}-c_{2j}\\ \frac{1}{2\pi}{\cF}_j
\end{array}\right)
\stackrel{\rm{T}}{\longrightarrow}\left(
\begin{array}{c}
\hat{h}_{1j}+\hat{c}_{1j}\\\hat{h}_{2j}+\hat{c}_{2j}\\\hat{h}_{1j}-\hat{c}_{1j}\\\hat{h}_{2j}-\hat{c}_{2j}  \\\frac{1}{2\pi} \hat{\cF}_j
\end{array}\right)
=
g\left(
\begin{array}{c}
{h}_{1j}+c_{1j}\\{h}_{2j}+c_{2j}\\{h}_{1j}-c_{1j}\\{h}_{2j}-c_{2j}\\\frac{1}{2\pi} {\cF}_j
\end{array}\right)\hsp g^\top \eta \, g=\eta \, , \label{th2}
\eeq 
where $c_{Ij} = c^I_{j}$.
One of our principal claims is that the normalization in the transformation (\ref{th2}) is correct. The last entry $\frac{1}{2\pi}{\cF}_j$ is always even as an element of integral cohomology, therefore we will need to demonstrate that the other entries are also even so that they may be rotated amongst each other by arbitrary integral orthogonal transformations. 

We will first demonstrate that, with the choice of $H$ in (\ref{hdj}), which is imposed by supersymmetry, the right-moving circle bundle curvature $h-c$ also satisfies the Hermitian Yang-Mills equations and so can be consistently exchanged with the $U(1)$ gauge bundle field strength.  We first evaluate $H$, by substituting (\ref{J}) into (\ref{hdj}).  We ignore the derivatives on $e^{2\phi} J_B$, as $\phi$ and $J_B$ may be pulled back from the base these derivatives may also be pulled back from the base and so will not contribute to $h$.  This leaves
\begin{eqnarray}
H&=&-\frac{1}{2}\overline{\partial}(\Theta\wedge\overline{\Theta})+\frac{1}{2}{\partial}(\Theta\wedge\overline{\Theta})\\&=&-\frac{1}{2}(\overline{\partial}\Theta)\wedge\overline{\Theta}+\frac{1}{2}\Theta\wedge\overline{(\partial\Theta)}+\frac{1}{2}(\partial\Theta)\wedge\overline{\Theta}-\frac{1}{2}\Theta\wedge\overline{(\overline{\partial}\Theta)}\nonumber\\
&=&-\frac{1}{2}\wa\wedge\overline{\Theta}+\half\Theta\wedge\overline{\ws}+\half\ws\wedge\overline{\Theta}-\half\Theta\wedge\overline{\wa}\nonumber\\
&=&\half(\ws-\wa)\wedge\overline{\Theta}+\half\Theta\wedge(\overline{\ws}-\overline{\wa}).
\end{eqnarray}

It is not $H$ that combines with the Chern classes in the T-duality doublet, but rather its pushforward $h=\pi_* H$ by the projection map $\pi$ of the circle bundle.  There are two circle bundles, one with fiber $\theta^1$ and one with fiber $\theta^2$ and correspondingly two projection maps, $\pi^1$ and $\pi^2$.  The two pushforwards act on the 1-forms as
\beq
\pi^1_*(\Theta)=\pi^1_*(d\theta)=\pi^1_*(d\theta^1)=1\hsp \pi^2_*(\Theta)=\pi^2_*(id\theta^2)=i
\eeq
and similarly the pushforwards on the conjugate $\overline{\Theta}$ yields the complex conjugates $1$ and $-i$.  On the other hand the pushforwards of $F$ are zero, and so only the pushforwards of the $\Theta$'s contribute to the pushforwards of $H$
\beq
h_1= \frac{1}{2\pi} \pi^1_*(H)=\Re(\ws-\wa)\hsp h_2=\frac{1}{2\pi} \pi^2_*(H)=\Im(\ws-\wa).
\eeq

These are very similar expressions to the Chern classes of the two circle bundles, which are
\beq
c_1^1=\frac{1}{2\pi} \Re(\ws+\wa)\hsp c_1^2=\frac{1}{2\pi} \Im(\ws+\wa).
\eeq
The expressions for $h$ and $c_1$ may be combined into left-moving and right-moving parts, which both lift to integral cohomology.  In particular the right-movers are
\beq
\frac{1}{2\pi}F_R^1=\half(h_1-c^1_1)=\frac{1}{2\pi}\Re\wa\hsp
\frac{1}{2\pi}F_R^2=\half(h_2-c^2_1)=\frac{1}{2\pi}\Im\wa.
\eeq
whereas the left movers are given by the same expressions, but using the self-dual part of the curvature.  The quantization of the real and imaginary parts of $\wa$ now imply that, as is required for the consistency of the T-duality (\ref{th2}), $h-c$ is even.

The consistency of the T-duality transformations which exchange the circle and gauge bundles demands, if the supersymmetry is to be manifest in the supergravity description, that $\wa$ solve the Hermitian Yang-Mills equations.  As $\wa$ is a $(1,1)$-form, it clearly satisfies the first two conditions.  Thus one need only show that its' contraction with $J$ vanishes
\beq
F_{R}^{ab}J_{ab}=\star(F_R\wedge J\wedge J)=\star(F_R\wedge(e^{4\phi}(J_B\wedge J_B)+ie^{2\phi}J_B\wedge\Theta\wedge\overline{\Theta})). \label{fj}
\eeq
Notice that each term on the right hand side vanishes separately using the primitivity condition (\ref{prim}).  Therefore the entire expression is zero and $F_R$ satisfies the Hermitian Yang-Mills equation.  

We must now the check the converse, that a solution to Hermitian Yang-Mills together with a supersymmetric solution for the right-movers yields a primitive circle bundle field strength and an $H$ flux that satisfies Eq.~(\ref{hdj}).  T-duality exchanges the Hermitian Yang-Mills curvature ${\cF}$ with the real or imaginary part of $F^-_{(1,1)}$, therefore we need to demonstrate that ${\cF}$ satisfies all of the conditions satisfied by $F^-_{(1,1)}$. In particular ${\cF}$ must be primitive on the base.

The crucial observation in this demonstration is that the toroidal components of the curvature ${\cF}$ vanish \cite{BY}.  This implies that the first term on the right hand side of Eq.~(\ref{fj}) vanishes, leaving
\beq
{\cF}^{ab} J_{ab}=ie^{2\phi}\star({\cF}\wedge J_B\wedge\Theta\wedge\overline\Theta).
\eeq
The left hand side vanishes because ${\cF}$ is Hermitian Yang-Mills.  On the right hand side, ${\cF}\wedge J_B$ may be pulled back from the base, whereas $\Theta\wedge\overline\Theta$ has both legs along the fiber and is nonzero.  Therefore the vanishing of the expression implies that ${\cF}\wedge J_B$ vanishes, otherwise its product with $\Theta\wedge\overline\Theta$ would be nonzero.  Thus ${\cF}\wedge J_B$ is primitive as desired, and so T-duality not only takes torus bundles with primitive curvatures to Hermitian Yang-Mills gauge bundles, but also takes Hermitian Yang-Mills gauge bundles to torus bundles with primitive curvatures.  Hence we have proven that T-duality preserves the supersymmetry of the low energy supergravity description, despite the fact that the low energy description is not valid as the torus area is of order $\alpha^\prime$.

\subsection{T-duality preserves the tadpole condition}

Now it is easy to show that this T-duality also preserves the tadpole condition.  According to Ref.~\cite{BY}, the tadpole condition on the K3 base is
\beq
\frac{1}{4\pi^2}\int_{K3} (4\ws\wedge\overline{\ws}-4\wa\wedge\overline{\wa}-\tr{\cF}^2)= 96. \label{tby}
\eeq
First consider the above T-duality transformation, which exchanges $\wa$ with ${\cF}/2$ while leaving $\ws$ constant.  By assumption, the gauge group is abelian, therefore $\tr({\cF}^2)={\cF}^2$ and so this T-duality transformation leaves the tadpole condition (\ref{tby}) fixed.

To see that the tadpole condition (\ref{tby}) is invariant under an arbitrary T-duality, notice that, multiplying by $4\pi^2$, it may be re-expressed as a product in integer cohomology
\beq
96= v^\perp \eta v\hsp
v= \frac{1}{2\pi} \left(
\begin{array}{c}
2F_{Lj}^1\\2F_{Lj}^2\\2F_{Rj}^1\\2F_{Rj}^2\\{\cF}_j
\end{array}\right)=
\left(
\begin{array}{c}
{h}_{1j}+c_{1j}\\{h}_{2j}+c_{2j}\\{h}_{1j}-c_{1j}\\{h}_{2j}-c_{2j}\\{\cF}_j
\end{array}\right)
\eeq
where $\eta$ is the diagonal unit metric with signature $(2,3)$.  In fact it is just the vanishing of the invariant $X_4$ as a 4-cohomology class on $B$.  This expression is manifestly $SO(2,3;\Z)$-invariant, because according Eq.~(\ref{th2}) $v$ transforms in the fundamental of $SO(2,3;\Z)$ and $SO(2,3;\Z)$ matrices by definition preserve the metric $\eta$.  Therefore the global tadpole condition is T-duality invariant with precisely the normalization of $\cF$ in Eq.~(\ref{th2}).

In fact  a local version of (\ref{tby}) holds, i.e  the integrand  is invariant under the action of $SO(2,3;\Z)$.  Then it follows from the unintegrated tadpole condition that
\beq
2i \partial \bar{\partial} e^{2\phi} \wedge J_B - \frac{\alpha'}{4} \tr R\wedge R 
\eeq
must also be T-duality invariant.

\section* {Acknowledgement}

We would like to thank P. Bouwknegt, V. Mathai and I. Melnikov for discussions and comments on the manuscript.  J.E. would like to acknowledge the Universit\'e Libre de Bruxelles for hosting him while some of this work was in progress.  R.M. would like to thank Max Planck Institute for gravitational physics at Potsdam  for hospitality and A. von Humboldt  foundation for support; R.M.
is supported in part by RTN contract  MRTN-CT-2004-005104 and
by ANR grant BLAN06-3-137168.

\appendix

\section{An example and K-theory}

\subsection{Narain T-duality over $S^2\times S^2$}

The simplest nontrivial example with a rank 2 $c^I{}_i$ matrix  is a $\mathbb{T}^2$ bundle $X$ over $S^2\times S^2$. Name the two circles $S^1_1$ and $S^1_2$ and the two 2-spheres $S^2_1$ and $S^2_2$.  Let the first cohomology group $\Z$ of $S^1_I$ be generated by the elements $\Theta^I$ and the let the second cohomology group $\Z$ of $S^2_i$ be generated by $\beta^i$.  The cup product of these two 2-classes is equal to the generator of the fourth cohomology of $S^2\times S^2$.  

The only relevant features of the topology of $S^2\times S^2$ for the present example are that it is simply connected, that the second Betti number is equal to two, and that the intersection matrix of the two two-cycles is the 2-dimensional off-diagonal matrix with entries equal to one.  Therefore, the following example may also be applied to any pair of 2-cycles in K3 with the same intersection number.  The generalization to general bundles on K3 will lead to different factors coming from the intersection matrix, and so the particular integers in this example will need to be modified.  However the process of T-dualizing a $\mathbb{T}^2$ fibered over K3, up to these intersection matrix factors, proceeds identically to the example of this section.

The third cohomology of the trivial bundle is $\Z^4$ and it is generated by a choice of $S^2$ and $S^1$
\beq
H^3(S^2_1\times S^2_2\times S^1_1\times S^1_2)=\Z^4=<\Theta^I\cup\beta^j>.
\eeq
Therefore an arbitrary $H$-flux may be, up to an exact form, decomposed as
\beq
H=2\pi \, h_{Ij}\Theta^I\cup\beta^j
\eeq
where Dirac quantization imposes that the matrix elements $h_{Ij}$ are integers.  Notice that in this example $H$ automatically has a single leg along the $\mathbb{T}^2$ as desired.

What if the $\mathbb{T}^2$ bundle is principal but nontrivial?  Then it will be characterized by the first Chern classes $c_1^i$ of the $S^1_i$ bundles, which are elements of $\H^2(S^2\times S^2)=\Z^2$.  We will decompose these in terms of the second cohomology as $c_1^I=c^I{}_{j}\beta^j$.

The $H$-flux may be expressed in the same basis, as the forms $\Theta^I$ and $\beta^j$ continue to exist in the nontrivial case. However the $\Theta^I$'s are no longer closed, instead
\beq
\frac{1}{2 \pi} d\Theta^I=c_1^I=c^I{}_{j}\beta^j.
\eeq
  This implies that in general the $H$-flux is also not closed 
\beq
dH= 2\pi \,h_{Ij}d\Theta^I\cup\beta^j=  h_{Ij}c^I{}_{k}\beta^k\cup\beta^j=h_{I1}c^I{}_{2}+h_{I2}c^I{}_{1} \label{ns5}
\eeq
where we have used the fact that the cup product of $S^2_i$ and $S^2_j$ is equal to one if each sphere appears once and we have expressed the final answer in units of the top form of $S^2\times S^2$.  The index $I$ is summed over.  

Again $c^I{}_j$ and $h_{Ij}$ are integers which entirely characterize the topology, and their contractions with $\beta^j$ lift to integral cohomology 2-classes.  In the rest of this appendix, the integral classes will be multiplied using the full integral cup product.

In summary, $\mathbb{T}^2$ fibrations over $S^2\times S^2$ with $H$-flux are characterized by 8 numbers, the two by two matrices $c_{ij}$ and $h_{ij}$.  These may be organized into two 2-vectors 
\beq
v_i=\left(\begin{array}{c}c^1_i\\c^2_{i}\\h_{1i}\\h_{2i}\end{array}\right)
\eeq
which transform in the fundamental of $O(2,2;\Z)$ in a basis in which elements $g$ of $O(2,2;\Z)$ satisfy Eq.~(\ref{conoluce}).  In this basis $O(2,2;\Z)$ matrices are those which preserve the metric $\eta$ of Eq.~(\ref{h}).
The $Q_{(5)}$ charge is just the inner product of $v_1$ and $v_2$ under the matrix $\eta$.  Therefore the full Narain T-duality group $O(2,2;\Z)$ preserves the charge.  In particular, a configuration with no charge will always be T-dual to a configuration with no charge.

For example, consider a $\mathbb{T}^2$ bundle $X$ with first Chern classes
\beq
c^1=(4,0)\in\H^2(S^2\times S^2)=\Z^2\hsp
c^2=(0,6)\in\H^2(S^2\times S^2)=\Z^2
\eeq
corresponding to the $c$-matrix
\beq
c^I{}_{j}=\left(\begin{array}{cc}4&0\\0&6\end{array}\right). \label{cex}
\eeq
The total space $X$ is then the product of lens spaces
\beq
P=L(4,1)\times L(6,1).
\eeq
The cohomology of $X$ may be found using the Gysin sequence, for example
\beq
\H^3(X)=\Z^2\oplus\Z_2. \label{3ex}
\eeq
The generators of the $\Z^2$ are $\Theta^I\cup\beta^i$ and the $\Z_2$ is generated by $-3\Theta^1\cup\beta^2+2\Theta^2\cup\beta^1$ which is $\Z_2$-nilpotent because when multiplied by 2 it becomes $d(\Theta^1\cup\Theta^2)$.  Any matrix $h$ which is not a linear combination of these generators will describe a configuration with $Q_{(5)}$ charge. 

Consider for example
\beq
H= 2\pi \, \Theta^1\cup\beta^1
\eeq
which corresponds to 
\beq
h=\left(\begin{array}{cc}1&0\\0&0\end{array}\right). \label{hex}
\eeq
This is a single unit of $H$-flux on the lens space $L(4,1)$.  $H$ is closed and so there are no NS5-branes.  Now we will investigate the actions of several elements $g$ of the Narain T-duality group $O(2,2;\Z)$.  The element
\beq
g=\left(\begin{array}{cccc}0&0&1&0\\0&1&0&0\\1&0&0&0\\0&0&0&1\end{array} \right)
\eeq
interchanges the Chern class of $S^1_1$ with $H$-flux with a leg along $\Theta^1$.  This element has determinant $-1$, and so interchanges the two twisted K-groups of $X$ as was shown in Ref.~\cite{BEM}.  While $S^2_2$ supports neither the Chern class of $S^1_1$ nor $H$-flux with a component $\Theta^1\cup\beta^2$, the transformation $g$ acts trivially on the lens space $L(6,1)$.  Instead it interchanges the $H$-flux and Chern class of $L(4,1)$, leaving an $S^3$ with $4$ units of $H$-flux.  Therefore $g$ represents an ordinary T-duality of $S^1_1$, which exchanges $L(4,1)\times L(6,1)$ with 1 unit of $H$ on $L(4,1)$ into $S^3\times L(6,1)$ with $4$ units of $H$ on $L(4,1)$.

The new T-dualities are transformations $g$ that mix the various circles.  For example, consider
\beq
g=\left(\begin{array}{cccc}-1&3&6&2\\0&1&2&0\\0&1&1&0\\1&-3&-3&-1\end{array} \right) \label{nontriv}
\eeq
acting on the configuration with Chern class (\ref{cex}) and $H$-flux (\ref{hex}).  This yields the dual configuration
\beq
\left(\begin{array}{c}c^I{}_{j}\\h_{Ij}\end{array}\right)=\left(\begin{array}{cccc}-1&3&6&2\\0&1&2&0\\0&1&1&0\\1&-3&-3&-1\end{array} \right)\left(\begin{array}{cc}4&0\\0&6\\1&0\\0&0\end{array}\right)=\left(\begin{array}{cc}2&18\\2&6\\1&6\\1&-18\end{array}\right).
\eeq
An $SL(2,\Z)$ rotation, subtracting $S^1_2$ from $S^1_1$, yields a slight simplification
\beq
\left(\begin{array}{c}c^I{}_{j}\\h_{Ij}\end{array}\right)=\left(\begin{array}{cc}0&12\\2&6\\1&6\\2&-12\end{array}\right).
\eeq
Notice that the two column vectors are still orthogonal with respect to the metric (\ref{h}), as is guaranteed by the fact that $g\in O(2,2;\Z)$, and the invariant $Q_{(5)}$ is unchanged.

\subsection{Twisted K-theory}

The determinant of $g$ in Eq.~(\ref{nontriv}) is equal to one, and so it preserves the theory and we may expect that it preserves both twisted K-groups of the compactification space.  To calculate these, we need to first use the Gysin sequence to find the cohomology of the $\mathbb{T}^2$ bundle $X$ and then we need to use the Atiyah-Hirzebruch spectral sequence to reduce the cohomology to the twisted K-theory.  The spectral sequence does not precisely produce the K-groups, instead it yields an associated graded group in which the additive structure of torsion subgroups may be somewhat different.  On the other hand it is the K-groups themselves which are preserved by T-duality, and so one may find that the T-dual associated graded K-groups have slightly different additive structures.

In the original fibration, the two lens spaces may be treated independently because both the Chern classes and the $H$ flux projections are supported on individual spheres.  Therefore one may find the twisted K-theories of both lens spaces and multiply them using the K\"unneth theorem.  The twisted K-groups of the two lens spaces are
\beq
\K^0_H(L(4,1))=\Z_4\hsp \K^1_H(L(4,1))=0\hsp \K^0_H(L(6,1))=\Z\oplus\Z_6\hsp \K^1_H(L(6,1))=\Z.
\eeq
The K\"unneth theorem then gives
\begin{eqnarray}
\K^0_H(L(4,1)\times L(6,1))&=&\Z_4\oplus{\textrm {Tor}}(\Z_4,\Z_6)=\Z_4\oplus\Z_{2}\nonumber\\
\K^1_H(L(4,1)\times L(6,1))&=&\Z_4\oplus\Z_4\otimes\Z_6=\Z_4\oplus\Z_{2}.
\end{eqnarray}

To find the twisted K-theory of the T-dual space, we first need to find its cohomology.  This is somewhat simplified by the fact that $S^1_1$ is only nontrivially fibered over $S^2_2$.  Therefore the $S^1_1$ fibration over $S^2\times S^2$ is $B=S^2_a\times L(12,1)$.  The cohomology of $B$ may be determined from the K\"unneth theorem, it is
\beq
\H^0(B)=\H^5(B)=\Z\hsp \H^1(B)=0\hsp \H^2(B)=\Z\oplus\Z_{12}\hsp \H^3(B)=\Z\hsp \H^4(B)=\Z_{12}.
\eeq

Now we need to fiber $S^1_2$ over $B$, which yields the T-dual total space $X$.  As $X$ is orientable
\beq
\H^0(X)=\H^6(X)=\Z.
\eeq
The nontriviality of $c^2$, the first Chern class of the $S^1_2$ bundle, together with the fact that $B$ is simply connected implies that the first homology of $X$ has no free part and so, by the universal coefficient theorem, the first cohomology is trivial
\beq
\H^1(X)=0.
\eeq
The Gysin sequence for the cohomology of $X$ begins with the short exact sequence
\beq
0\stackrel{\pi_*}{\longrightarrow}\H^0(B)=\Z\stackrel{c^2\cup}{\longrightarrow}\H^2(B)=\Z\oplus\Z_{12}\stackrel{\pi^*}{\longrightarrow}\H^2(X)\stackrel{\pi_*}{\longrightarrow}0
\eeq
which expresses the second cohomology group of $X$ as the quotient
\beq
\H^2(X)=\frac{\H^2(B)}{\H^0(B)}=\frac{\Z\oplus\Z_{12}}{\Z=<2,6>}
\eeq
where the notation $<2,6>$ refers to the $\Z$ generated by the element $(2,6)\in\Z\oplus\Z_{12}$, which is the Chern class.  This quotient is $\Z_{12}\oplus\Z_2$, where the $\Z_{12}$ is generated, for example, by $(0,1)$ and the $\Z_2$ by $(1,3)$.  Therefore
\beq
\H^2(X)=\Z_{12}\oplus\Z_2.
\eeq

To find the third and fourth cohomology groups we need to use the later part of the sequence
\begin{eqnarray}
&&0\stackrel{c^2\cup}{\longrightarrow}\H^3(B)=\Z\stackrel{\pi^*}{\longrightarrow}\H^3(X)\stackrel{\pi_*}{\longrightarrow}\H^2(B)=\Z\oplus\Z_{12}\stackrel{c^2\cup}{\longrightarrow}\H^4(B)=\Z_{12}\stackrel{\pi^*}{\longrightarrow}\\&&\stackrel{\pi^*}{\longrightarrow}\H^4(X)\stackrel{\pi_*}{\longrightarrow}\H^3(B)=\Z\stackrel{c^2\cup}{\longrightarrow}\H^5(B)=\Z\stackrel{\pi^*}{\longrightarrow}\H^5(X)\stackrel{\pi_*}{\longrightarrow}\H^4(B)=\Z_{12}\stackrel{c^2\cup}{\longrightarrow}0.\nonumber
\end{eqnarray}
Tensoring the sequence by $\R$, one finds that $\H^3(X)$ is in a short exact sequence between two $\R$'s and so its free part is of dimension two.  Therefore $\H^3(X)$ is equal to $\Z^2$ plus a torsion term.  As $\H^3(B)$ is torsion free, no torsion comes from the left and so the $\pi_*$ must take this torsion injectively into $\H^2(B)=\Z\oplus\Z_{12}$.  Therefore the torsion term is just the torsion part of the kernel of $c^2$ cupping $\H^2(B)$.  Given an element
\beq
(a,b)\in\Z\oplus\Z_{12}=\H^2(B)
\eeq
the cup product with $c^2$ is
\beq
c^2\cup(a,b)=(2,6)\cup(a,b)=6a+2b\in\Z_{12}=\H^4(B).
\eeq
The kernel of this map is $\Z\oplus\Z_2$, where the $\Z$ is generated, for example, by $(a,b)=(1,3)$ and the $\Z_2$ by $(0,6)$.  The torsion part of the kernel is $\Z_2$, which we have argued came from $\H^3(X)$ and so
\beq
\H^3(X)=\Z^2\oplus\Z_2=<\Theta^1\cup\beta^2,\Theta^2\cup\beta^1-3\Theta^2\cup\beta^2,6\Theta^2\cup\beta^2-\Theta^1\cup\beta^1-3\Theta^1\cup\beta^2>. \label{h3}
\eeq

By Poincar\'e duality $\H_3(X)=\Z^2\oplus\Z_2$ and so by the universal coefficient theorem the fourth cohomology group has the free part of $\H^2$, which is trivial, and a torsion subgroup $\Z_2$, therefore
\beq
\H^4(X)=\Z_2.
\eeq
Similarly the universal coefficient theorem yields $\H_1(B)=\Z_{12}\oplus\Z_{2}$ and so by Poincar\'e duality
\beq
\H^5(X)=\Z_{12}\oplus\Z_2
\eeq
completing the calculation of the cohomology of $X$.

To find the associated graded form of the twisted K-theory of $X$ we need only take the cohomology of $\H^*(X)$ with respect to the operator $H\cup$, there are no other contributions to the twisted K-theory in the case of a simply-connected, oriented 6-manifold \cite{0804.0750}.  The $H$ flux is
\beq
\frac{1}{2\pi} H=\Theta^1\cup\beta^1+6\Theta^1\cup\beta^2+2\Theta^2\cup\beta^1-12\Theta^2\cup\beta^2.
\eeq

Let us begin with $K^0_H(X)$, we would like this to be $\Z_4\oplus\Z_2$, or at least to have 8 elements, like the original $K^0$.  It includes $H^6(X)$ quotiented by $H\cup\H^3(X)$.  The only nonzero terms in this cup product come from the free part of $\H^3(X)$, which is $\Z^2$.  One $\Z$ came from $\H^3(L(12,1))$ and is generated by $\Theta^1\cup\beta^2$.  $H\cup$ this term yields the volume form with weight two.  The other $\Z$ is generated by $\Theta^2\cup(\beta^1-3\beta^2)$ and when cupped with $H$ yields $-3$ times the volume form.  Therefore all of $\H^6$ is in the image of $H\cup$ and so $\K^0_H(X)$ will consist only of those elements of $\H^4(X)$ and $\H^2(X)$ which are not in the image of $H\cup$.  No element of $\H^2(X)$ may be in the image of $H\cup$, as the image is of rank at least 3.  Similarly no element of $\H^4(X)$ is in the image, because it would need to be $H$ cupped with a 1-class, but $\H^1(X)=0$.  However only those elements in the kernel of $H\cup$ contribute to the twisted K-theory.  All elements of degree greater than $3$ are in the kernel, no elements at degree 0 are, and so $\K^0$ will be the sum of $\H^4(X)$ and those elements of $\H^2(X)$ which are annihilated by $H\cup$.

We then need to know the action of $H\cup$ on $\H^2(X)$.  $\H^2(X)$ is the sum of $\Z_{12}$ and $\Z_2$ which are generated by $\beta^2$ and $\beta^1+3\beta^2$ respectively.  The image is in $\H^5(X)$ which is also a sum of $\Z_{12}$ and $\Z_2$, generated by $\Theta^2\cup\beta^1\cup\beta^2$ and $\Theta^1\cup\beta^1\cup\beta^2$ respectively.  The generator of the $\Z_{12}$ appears to go to $(2,1)\in\Z_{12}\oplus\Z_2$, and the generator of $\Z_2$ to $(6,1)$, which is just 3 times the image of the generator of $\Z_{12}$. Therefore the image of $\H^2(X)$ consists of $6$ elements of the 24 in $\Z_{12}\times\Z_2$, which are the 6 multiples of $(2,1)$.  The kernel therefore consists of $24/6=4$ elements.

Therefore, at least as a set
\beq
\K^0_H(X)=\Z_4\oplus\H^4(X)=\Z_4\oplus\Z_2.
\eeq
As a group this is  not necessarily equivalent to the original $\K^0$, but this is to be expected from an associated graded part.  However both sets have 8 elements, which is supporting evidence for the conjecture that the K-theories are isomorphic.  

Next we turn our attention to $\K^1$, which had order 8 before the T-duality.  As $\H^1(X)=0$, the contributions will be the elements of $\H^5(X)$ modulo those which are $H\cup$-exact, and the $H\cup$-closed elements of $\H^3(X)$ modulo the exact.  We have already seen that the image of $H\cup$ in $\H^5(X)$ contains 6 of the 24 elements, therefore $\H^5(X)$ contributes $24/6=4$ elements to $\K^1_H(X)$.  Next we need to find the $H\cup$-closed elements of $\H^3(X)$.  In the basis of Eq.~(\ref{h3}) these are elements of the form $(3a,2a,a+b)$ where the last entry is the $\Z_2$ and $a$ and $b$ are arbitrary integers.  In this basis
\beq
H=(3,2,-1)
\eeq
and so the $a$ parameter labels the $H\cup$-exact part, and the $\Z_2$-valued $b$ parameter contributes a $\Z_2$ to $\K^1$.  Therefore, as a set $\K^1$ contains a $\Z_4$ from $\H^5$ and a $\Z_2$ from $\H^3$
\beq
\K^1(X)=\Z_4\oplus\Z_2.
\eeq
Thus it has again 8 elements, as was the case before the T-duality.  Unfortunately as $\K^0$ and $\K^1$ have the same number of elements in this example, it is impossible to determine whether or not they were exchanged by the T-duality.


\end{document}